\documentclass{llncs}
\usepackage{graphicx}
\usepackage[usenames,dvipsnames]{xcolor}
\usepackage{amsmath}
\usepackage{amsfonts}
\usepackage[utf8]{inputenc}
\usepackage{listings}
\usepackage{booktabs}
\usepackage{graphics}
\usepackage{wrapfig}
\usepackage[
        backend=bibtex,
        url=false,
        ]{biblatex}

\DeclareFieldFormat{title}{\mkbibemph{\href{\thefield{url}}{\thefield{title}}}}

\begin{document}

\title{Mathoid:~Robust, Scalable, Fast and Accessible Math Rendering for Wikipedia}

\author{Moritz Schubotz\inst{1}, Gabriel Wicke\inst{2}}
\institute{Database Systems and Information Management Group,\\
Technische Universit\"{a}t Berlin,
Einsteinufer 17, 10587 Berlin, Germany\\
\email{schubotz@tu-berlin.de}
\and Wikimedia Foundation, San Francisco, California, U.S.A.\\
\email{gwicke@wikimedia.org
}}

\maketitle
\setcounter{footnote}{0}
\begin{abstract}
Wikipedia is the first address for scientists who want to recap basic mathematical 
and physical laws and concepts.
Today, formulae in those pages are displayed as Portable Network Graphics images. 
Those images do not integrate well into the text, can not be edited after copying, 
are inaccessible to screen readers for people with special needs,
do not support line breaks for small screens and do not 
scale for high resolution devices.
Mathoid improves this situation and converts formulae specified by Wikipedia editors in 
a \TeX-like input format to MathML, with Scalable Vector Graphics images as a fallback
solution.
\end{abstract}

\author{Moritz Schubotz, Gabriel Wicke} 

\section{Introduction: browsers are becoming smarter}
\label{sc.intro}

Wikipedia has supported mathematical content since 2003.
Formulae are entered in a \TeX-like notation 
and rendered by a program called {\tt texvc}. One of the first 
versions of {\tt texvc} announced the future of MathML support as 
follows: 
\begin{quotation}
\noindent ``As of January 2003, we have TeX markup for mathematical formulas on Wikipedia.
It generates either PNG images or simple HTML markup, depending on user prefs and the 
complexity of the expression.
In the future, as more browsers are smarter, 
it will be able to generate enhanced HTML or even MathML in many cases.''
\cite{texvcHelp2003}
\end{quotation}
Today, more then 10 years later, less than 20\% of people visiting the 
English Wikipedia, currently use browsers that support MathML (e.g., Firefox) \cite{Zachte2013}.
In addition, {\tt texvc}, the program that controls math 
rendering, has made little progress in supporting MathML. 
Even in 2011, the MathML support was ``rather pathetic'' (see \cite{MathMLBroken}).
As a result, users expected MathML support within Wikipedia to be a broken feature.
Ultimately, on November 28, 2011, the user preference for 
displaying MathML was removed \cite{MathMLSettingsRemoved}.

Annoyed by the Portable Network Graphics (PNG) images,
in December 2010, user Nageh published a script, \texttt{User:Nageh/mathJax.js}, 
that enables client-side MathJax rendering for individual Wikipedia users
\cite{userScriptToUseMathJax}.
Some effort and technical expertise was required to use the script.
The user had to install additional fonts on his system manually, to import the script, 
into his Wikipedia account settings
and to change the Math setting in his Wikipedia user account page to ``Leave it as \TeX''.
With client-side MathJax rendering, the visitor was able to choose from the context 
menu of each equation with the PNG image being replaced by either: 
(1) a Scalable Vector Graphics (SVG) image, 
(2) an equivalent HTML + CSS representation, or 
(3) MathML markup (this requires a MathML capable browser).

MathJax needs a significant amount of time to replace the \TeX-code on
the page with the above replacements. This amount of time is dependent
on the operating system, browser, and hardware configuration.  
For instance, we measured 133.06 s to load the page 
\href{https://en.wikipedia.org/wiki/Fourier_transform}{\em Fourier transform} 
in client side MathJax mode, as compared to 14.05 s for the page loading without math 
rendering (and 42.9 s with PNG images) on a typical Windows laptop with 
Firefox.\footnote{The measurement was done on a Lenovo T420 Laptop with the following hardware:
8GB RAM, 500GB HDD, CPU Intel Core i7-2640M, Firefox 24.0 on Windows 8.1, 
download speed 98.7 MB/s upload speed 9.8 MB/s, ping to {\tt en.wikpedia.org} was $25(\pm1)$ ms.} 
However, improvements in the layout motivated many users to use that script, and
in November 2011, client-side MathJax became an experimental option
for registered users \cite{MathJax2mweMath}.

As of today, there are almost 21M registered Wikipedia users,
of which 130k have been active in the last 30 days.
Of these users, 7.8k use the MathJax rendering option which causes 
long waiting times for pages with many equations.
Also 47.6k users chose the (currently disabled) HTML rendering mode,
which if possible, tries to use HTML markup to display formula, and the 
PNG image otherwise.  Furthermore, 10.1k users chose the MathML 
rendering mode (disabled in 2011).
Thus, the latter 57.7k users are temporarily forced to view PNG images, 
even though they explicitly requested against this. 
This demonstrates that there is an significant demand for math rendering,
other than for the use of PNG images.

Currently, the MediaWiki Math extension is version 1.0.  Our efforts have
been to make an improvement on that extension.  We refer to our update of
the extension as version 2.0.  Furthermore, we refer to Mathoid as all the
ingredients mentioned in this paper which go into developing Math 2.0.
One essential ingredient in Mathoid, is what we refer to as Mathoid-server.  
This is a tool, which we describe in this paper, which converts the \TeX-like 
math input used in MediaWiki to various formats that we describe in this paper.
Our paper is organized as follows.

In Section \ref{sc.req}, we list the requirements for Math rendering in Wikipedia, 
explain how one may contribute to these requirements,
and elaborate on how one may make math accessible for people with special needs. 
In this section we introduce the Mathoid-server.
In Section \ref{sc.brws}, we explain how by using Mathoid-server, math can be 
displayed in browsers that do not support MathML.  
In Section \ref{sc.srv}, we discuss how math rendering can be offered
as a globally distributed service.
In Section \ref{sc.prfm}, we discuss and compare the performance of 
reviewed rendering tools, in regard to layout and speed.
Finally, in Section \ref{sc.cncl}, we 
conclude with results from our comparison and give an overview of future work.\\[-0.7cm]

\section{Bringing MathML to Wikipedia}
\label{sc.req}
\vspace{-0.2cm}
For Wikipedia, the following requirements and performance measures are critical.
\begin{description}
\item[coverage:] The converter must support all commands currently used in Wikipedia.
\item[scalability:] The load for the converter may vary significantly, 
since the number of Wikipedia edits heavily depends on the language.
Thus, a converter must be applicable for both small and large Wikipedia instances.
\item[robustness:] Bad user input, or a large number of concurrent requests, must 
not lead to permanent failure for the converter.
\item[speed:] Fast conversion is desirable for a good user experience.
\item[maintainability:] A new tool for a global site the size 
of Wikipedia must be able to handle tasks with a large management overhead.
Therefore, active development over a long period of time is desirable.
\item[accessibility:] Providing accessible content to everyone is one of the key goals of Wikipedia.
\end{description}

There are a large variety of \TeX~to MathML converters \cite{Bos}.
However, most of them are no longer under active development,
 or their licenses are not compatible with MediaWiki.
In 2009, \cite{StaGinDav:maacl09} showed that \LaTeX ML has the best coverage (but not a very high 
conversion speed) as compared to the \LaTeX~converters which were analysed in that paper.
Since 2009, a new converter, MathJax~\cite{Cervone}, has become popular.
After negotiations with Peter Krautzberger (of MathJax) and his team, we regard MathJax 
(executed in a headless browser on a web server), as a useful alternative to \LaTeX ML.
One strong point about MathJax with regard to coverage, is that it is already used by some Wikipedia 
users on the client-side (as described in Section \ref{sc.intro}).
Therefore the risk of unexpected behavior is limited.
For \LaTeX ML, Bruce Miller has written a set of custom macros for MediaWiki 
specific \TeX~commands which supports the MediaWiki \TeX~markup.
A test using these macros based on a Wikipedia dump, has shown very good coverage of the 
mathematics commands currently used in Wikipedia.
\LaTeX ML is maintained by the United States Department of Commerce Laboratory, the 
National Institute of Standards and Technology (NIST) and 
MathJax is maintained by the MathJax project which is a consortium of the 
American Mathematical Society and the Society for Industrial and Applied Mathematics.
We analyze and compare MathJax and \LaTeX ML in detail, since these are the most promising 
tools we could discover.

In regard to accessibility, we note that Wikipedia has recently made serious 
efforts to make the site 
more \href{http://en.wikipedia.org/wiki/Wikipedia:Accessibility}{accessible}.
However, images which represent equations are currently not accessible.
The only available information from PNG images (which is not very helpful) is 
the alt-text of the image that contains the \TeX~code.
Based upon recent unpublished work of Volker Sorge \cite{Sorge2014}, 
we would like to
provide meaningful semantic information for the equations.
By providing this, more people will be able to 
understand the (openly accessible) content \cite{AccessMathWeb}.
One must also consider that there is a large variety of special needs. 
People with blindness are only a small fraction of the target group which can benefit from 
increased accessible mathematical content.
Between 1 and 33\% \cite{Crystal1987,Czepita2006}
of the population suffer from dyslexia.
Even if we calculate with the lower boundary of 1\%,
90,000 people per hour visit the English Wikipedia and some of them
could benefit from improvements of the accessibility while reading articles that contain math.
However, our the main goal with regard to accessibility is to make Wikipedia accessible for blind people that 
have no opportunity to read the formulae in Wikipedia today. 

Furthermore, the information provided in the tree structure of mathematics
by using MathML, helps one to orientate complex mathematical equations,
which is useful for general purpose use as well.
With regard to accessibility, a screen reader can repeat only part of an equation to provide 
details that were not understood.
PNG images do not give screen readers detailed related mathematical 
information \cite{Chisholm2001,Miner2005}.
In 2007, \cite{Maddox2007} states that MathML is optimal for screen readers.
\href{http://www.washington.edu/doit/Faculty/articles?404}{\textit{The Faculty Room}}\footnote{\url{http://www.washington.edu/doit/Faculty/articles?404}}  website, 
lists four screen readers that can be used in combination with MathPlayer \cite{Soiffer2005} to read Mathematical equations.
Thus Mathoid-server and \LaTeX ML server~\cite{Ginev2011} that generate MathML output, contribute towards 
better accessibility within the English Wikipedia.\\[-0.7cm]

\section{Making math accessible to MathML disabled browsers}
\label{sc.brws}

\vspace{-0.1cm}
For people with browsers that do not support MathML, we would like to provide high 
quality images as a fallback solution. 
Both \LaTeX ML and MathJax provide options to produce images.
\LaTeX ML supports PNG images only, which tend to look rasterized if they are viewed 
using large screens.  
MathJax produces scalable SVG images.
For high traffic sites like Wikipedia with 9 million visitors per hour, it is crucial to reduce the server load generated by each visitor. Therefore rendered pages should be used for multiple visitors. 
Rendering of math elements is especially costly. 
This is related to the nested structure of mathematical expressions.
As a result, we have taken care that the 
output of the MediaWiki Math extension should be browser independent.
Since MathJax was designed for client-side rendering, our goal is to develop 
a new component. We call this new component the Mathoid-server. 
Mathoid-server, a tool written in JavaScript, uses MathJax to 
convert math input to SVG images. 
It is based on {\tt svgtex} \cite{svgtex} which uses {\tt nodejs} and {\tt phantomjs} to run MathJax 
in a headless browser. It exports SVG images.
Mathoid-server improves upon the functionality of {\tt svgtex} while offering a more
robust alternative.
For instance, it provides a restful interface which supports 
{\tt json} input and output as well as the support of MathML output.
Furthermore, Mathoid-server is shipped as a Debian package for easy 
installation and maintenance.
Many new developments in MediaWiki use JavaScript.
This increases the probability of finding volunteers to maintain the code and to fix bugs.
For general purpose, Mathoid-server can be used as 
a stand-alone service which can be used in other content management platforms 
such as Drupal or Wordpress.
This implies that Mathoid-server will have a larger interest group for maintenance in the future.
The fact that Mathoid-server automatically adapts to the number of 
available processing cores, and can be installed fully unattended via tools like Puppet,
indicates that the administrative overhead for Mathoid-server instances should be 
independent of the number of machines used.
In the latest version, the Mathoid-server supports both \LaTeX~and 
MathML input and is capable of producing MathML and SVG output.

To support MathML disabled browsers, we deliver both MathML markup, and a link to 
the SVG fallback image, to the visitor's browser. In order to be compatible with 
browsers that do not support SVG images, in addition, we add links to the old PNG images. 
In the future those browsers will disappear and this feature will be removed.

To prevent users from seeing both rendering outputs, the MathML element is hidden by default,
and the image is displayed.  For Mozilla based browsers (these support MathML rendering), we 
invert the visibility by using a custom CSS style, hide the fallback images and 
display the MathML-markup.  This has several advantages.
First, no browser detection, neither on the client-side (e.g., via JavaScript) nor 
on server-side is required. This eliminates a potential source of errors.
Our experiments with client-side browser detection showed that the user will 
observe a change in the Math elements if pages with many equations are loaded.
Second, since the MathML element is always available on the client-side, the user can 
copy equations from the page, and edit it visually with tools such as Mathematica.
If the page content is saved to disk, all information is preserved without resolving 
links to images.  If afterwards the saved file is opened with a MathML enabled browser, 
the equations can be viewed off-line.  This feature is less relevant for users 
with continuous network connections or with high-end hardware and software.
However, for people with limited resources and unstable connections 
(like in some parts of India~\cite{Arunachalam2006}), they will experience a significant benefit.

The current Firefox mobile version (28.0) passes the MathML Acid-2 test, indicating
that there is generally good support for MathML on mobile devices. 
This allows for customized high quality adaptable rendering for specific device properties.
The \href{http://www.w3.org/TR/MathML/chapter3.html#presm.linebreaking}{W3C MathML specification}\footnote{\url{http://www.w3.org/TR/MathML/chapter3.html}} 
discusses the so called best-fit algorithm for line breaking. According to our experiments,
Firefox-mobile (28.0) does not pass the W3C line break tests.
However, as soon this issue is fixed, mobile users will benefit 
from the adjusted rendering for their devices.  Note that there is 
active development in this area 
by the \href{http://www.ulule.com/mathematics-ebooks}{Mathematics in ebooks project}\footnote{\url{http://www.ulule.com/mathematics-ebooks}} 
lead by Fr\'{e}d\'{e}ric Wang.\\[-0.8cm]

\section{A global distributed service for math rendering}

\label{sc.srv}
\vspace{-0.1cm}
\begin{figure}[ht]
\centering
        
        \vspace{-10pt}
  \includegraphics[trim=55 55 55 15,clip,width=0.75\textwidth]{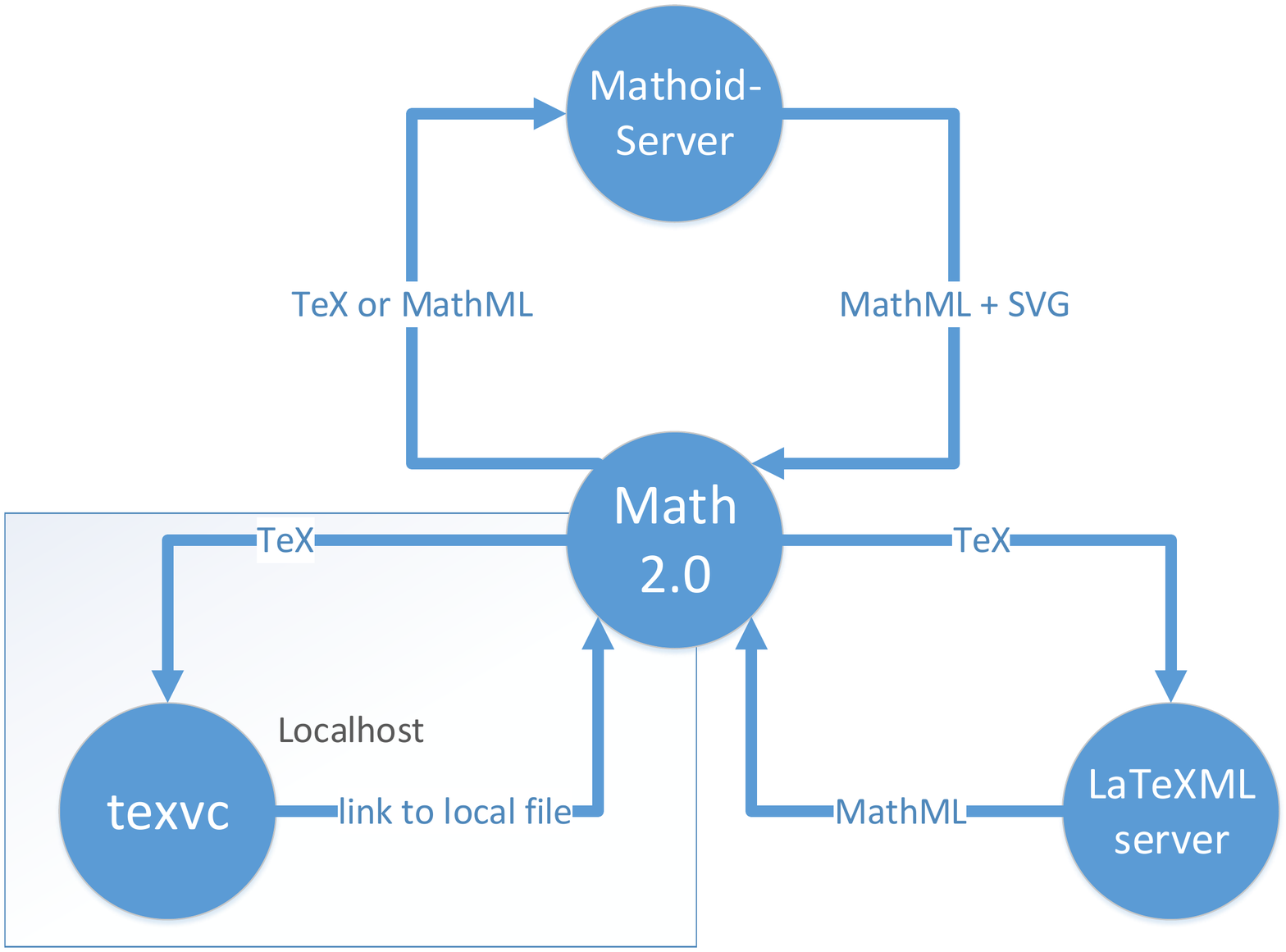}

\caption{System overview. The 2.0 release of the MediaWiki Math extension offers new ways 
to render Math input in MediaWiki. 
It communicates to \LaTeX ML servers and to instances 
of our MathJax-based development of the Mathoid-server. 
Note that we preserve backwards compatibility to the MediaWiki Math extension 1.0.}
  \label{fig.overview}
  \vspace{-10pt}
\end{figure}

To use \LaTeX ML and Mathoid-server for Math rendering within Wikipedia,
we have changed the MediaWiki Math extension (see Fig. \ref{fig.overview}).
While preserving backwards compatibility, we pursue our current 
development \cite{Schubotz} by integrating \LaTeX ML and Mathoid-server.
System administrators can choose which rendering back-ends are selectable in 
a MediaWiki instance by logged in users.  All rendering modes can be active at 
the same time, and it is possible to use Mathoid-server to generate SVG images based 
on the output of the \LaTeX ML server.

For {\tt texvc}, rendering requires one to install a 
full \LaTeX~distribution (about 1GB) on each web server. 
This is a huge administrative 
overhead and the management of files and 
permissions has caused a lot of problems. 
These problems were hard to solve and resulted in inconsistent 
behavior of the website from a user 
perspective~\cite{mediawiki:MathTalk,mediawiki:MathErrors,Bug54367,Bug54456}.
Furthermore, for MediaWiki instances run by individuals, it is difficult to set up math support. 
Therefore, one of our major requirements is that future versions of the MediaWiki 
Math extension should not need to access the file system.
Also, one should not need to use shell commands to directly access the server.
This has the potential to introduce major security risks.
With our separation approach, rendering and displaying of mathematics 
no longer needs to be done on the same machine.
However, for instances with a small load, this would still be possible. 
Furthermore small MediaWiki instances can now enable Math support without 
any configuration or shell access. By default, public \LaTeX ML and Mathoid-server 
instances are used. These will be provided by \href{https://www.xsede.org}{XSEDE}\footnote{\url{https://www.xsede.org}}. With this method, no additional security risk is provided 
for individuals. For Mathoid-server, the security risk for the host is limited as 
well. This is because the Mathoid process runs on a headless browser on the server 
without direct access to the file system.  

\paragraph{Caching.} There are two main caching layers.
In the first layer, the database caches the rendering result of the math conversion,
i.e., the MathML and SVG output in the 
database.\footnote{To keep the lookup time for equations constant, the key for the cache entry is a hash of \TeX~input.}
The second caching layer is browser based.
Similar to the SVG rendering process for ordinary SVG images, the MediaWiki Math extension uses cacheable special pages to deliver SVG images.
On the server side, those pages are cached by squid servers. 
In addition, even if images occur on different pages, the browser will only load 
that image once.\\[-0.7cm]
\section{Performance analysis}
\label{sc.prfm}
\begin{figure}[hp]
\centering
\includegraphics[trim=37 125 37 42,clip,width=1\textwidth,
]{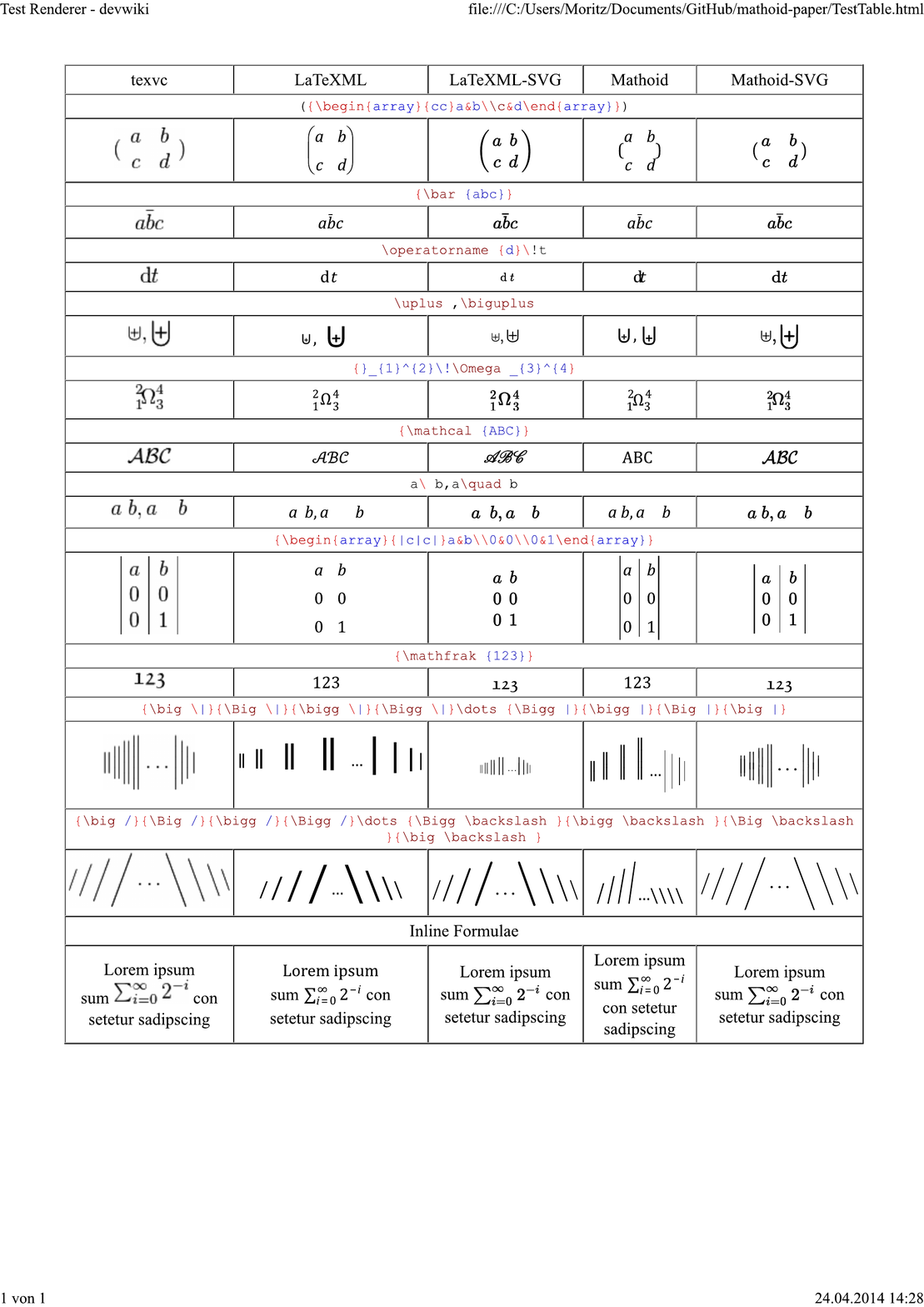}
\caption{This figure displays a comparison of possible rendering outputs for the MediaWiki Math 
extension rendered with Firefox 24.0.
Mathoid-server allows one to use either a \LaTeX ML or MathJax renderer to generate MathML 
output with SVG fallback.
The investigation of the listed corner cases shows that the Mathoid-SVG rendering option, that uses server-side MathJax rendering via {\tt phantomjs}, produces the best results.
\newline
{\small \it Remark: The authors thank Bruce Miller. He improved the \LaTeX ML implementation based on a preprint version of this paper. This final version of the paper uses the \LaTeX ML version of 24th of April 2014.}}
\label{fig.visComp}
\end{figure}

As a first step towards a visual analysis, we compared our impressions
of output from \LaTeX ML and MathJax using Firefox 24.0.
Except for additional {\tt mrow} elements in \LaTeX ML, the produced presentation MathML is 
almost identical.
The differences that we did notice had no influence on the rendering output.
In rare cases, MathJax uses {\tt mi} elements, whereas \LaTeX ML uses {\tt mo} elements.
In contrast to \LaTeX ML which uses UTF-8 characters to represent special symbols, 
MathJax uses HTML entities. 
However, there still remain some minor differences
(see Fig. \ref{fig.visComp}). 

To illustrate performance differences, we chose a random sample equation, namely
\begin{align}
{\frac {\left(x-h\right)^{2}}{a^{2}}}-{\frac {\left(y-k\right)^{2}}{b^{2}}}=1.
\end{align}
With 10 subsequent runs\footnote{All measurements were performed using 
virtual Wikimedia labs instances, with the following hardware specifications: number of CPUs 2 ,
RAM size 4096 Mb, allocated Storage 40 Gb}, the following conversion times were measured:
\begin{itemize}
\item \LaTeX ML : \TeX~$\to$ MathML (319ms/220ms);
\item Mathoid-server : \TeX~$\to$ SVG$\,+\,$MathML (18ms/18ms);
\item {\tt texvc} : \TeX~$\to$ PNG (99ms/71ms).
\end{itemize}
Thus compared to the baseline ({\tt texvc}), Mathoid-server produced a speedup of 5.5 
whereas \LaTeX ML is 3.2 times slower. \LaTeX ML and PNG seem to benefit from 
multiple runs, whereas the rendering time with Mathoid-server stays constant. 

We also converted all of the English Wikipedia articles and measured the conversion times 
for each equation therein.
The most time consuming equation was the full form of the \href{http://en.wikipedia.org/wiki/Ackermann_function}
{Ackermann function} $A(4,3)$.
\LaTeX ML\footnote{{\tt ltxpsgi} (\LaTeX ML version 0.7.99; revision ef040f5)} 
needs 147 s to answer the HTTP request for $A(4,3)$.
According to the self reported \LaTeX ML-log, approximately 136.18 s was used for parsing.
The same request was answered by Mathoid-server in 0.393 s,
which is approximately 374 times faster.
The old rendering engine needed 1.598 s to produce the image.
This does not include the 41 ms it took to load the image from the server.

\section{Conclusion, outlook and future work}
\label{sc.cncl}
\vspace{-0.1cm}

In the scope of Mathoid, we updated the retrograde MediaWiki Math extension, and 
developed Mathoid-server which replaces the {\tt texvc} program.
This enhanced the security of MediaWiki as proposed in \cite{Schubotz}.
It is no longer necessary to pass user input via shell access using the 
command-line. Nor is it necessary to move files on the server via PHP.
The MediaWiki Math extension is now capable of using \LaTeX ML and 
Mathoid-server to convert \TeX-like input to MathML$\,+\,$SVG.
Based on the requirements, the user can choose if he prefers to use \LaTeX ML for 
the MathML generation (this has the advantage of content MathML output), or he can use  
Mathoid-server which is much faster (but does not produce content MathML).
Mathoid-server takes advantage of \LaTeX ML since it produces MathML.
The MediaWiki math extension, through Mathoid-server, converts MathML to 
fallback SVG images.\footnote{This integral feature of Math 2.0 does not require additional 
source modifications, and is demonstrated for example at 
\href{http://demo.formulasearchengine.com}{\url{http://demo.formulasearchengine.com}}.}
For Wikipedia itself, with only a few semantic macros, and no real applications for 
content MathML produced by \LaTeX ML, Mathoid-server alone seems to be the best choice.

 \begin{table}
\centering
  
  \caption{Overview: comparison of different math rendering engines. \newline
     Values based on 
articles containing mathematics in the English Wikipedia.}
 \begin{tabular}{|l|c|c|c|}
 \hline  & {\tt texvc} & \LaTeX ML & Mathoid  \\ 
 \hline relative speed & 1 & 0.3 & 5\\
 \hline
 image output & PNG & PNG & SVG \\     
 \hline 
 presentation MathML coverage & low & high & high \\
 \hline
 content MathML output & no & no & yes \\
 \hline
  webservice & no & yes & yes \\
  \hline
  approximate space required on webserver & ~ 1GB\hspace{0.2cm} & 0 & 0 \\
 \hline
   language & OCaml\hspace{0.05cm} & Perl & JavaScript \\
  \hline
     maintained by & nobody & NIST & MathJax \\
    \hline
 \end{tabular} 
 \label{tb.overview}
 \end{table}
 
We did exhaustive tests to demonstrate the power of Mathoid-server 
with regard to scalability, 
performance and enhancement of user experiences. Those test results are summarized 
in Table \ref{tb.overview}. Our implementation was finished in October 2013 and is 
currently being reviewed by the Wikimedia foundation for production use.
Our work on the MediaWiki Math extension and Mathoid-server establishes a basis for 
further math related developments within the Wikimedia platform. Those developments might be 
realized by individuals or research institutions in the future.
For example, we have considered creating an OpenMath content dictionary~\cite{riem2004openmath}
that is based on Wikidata items~\cite{vrandevcic2012wikidata}.
This will augment mathematical formulae with language independent semantic information.
Moreover, one can use gathered statistics about formulae currently in use in
Wikipedia to \href{http://www.formulasearchengine.com/node/189}{enhance the user interface for entering new formulae}\footnote{\url{http://www.formulasearchengine.com/node/189}}. This is common for text input associated with mobile devices.

In regard to future potential work, one may note the following. 
There is a significant amount of hidden math markup in Wikipedia. 
Many of these have been entered using HTML workarounds (like subscript or superscript)
or by using UTF-8 characters. This markup is difficult to find and causes problems for 
screen readers (since they are not aware that the mode needs to be changed).  
If desired by the community, by using gathered statistics and edit history, we would be 
able to repair these damages.

The MediaWiki Math extension currently allows limited semantic macros. For example, one can 
use \texttt{\textbackslash Reals} to obtain the symbol $\mathbb{R}$. At the moment, 
semantic macros are seldomly used in the English Wikipedia (only 17 times). 
One could potentially benefit by increased use of semantic macros by taking advantage
of the semantic information associated with these.
In the future, one could use this semantic information to take advantage of additional 
services, such as MathSearch \cite{Schubotz}, a global search engine which
takes advantage of semantic information.  Also, the use of semantic macros would provide
semantic information, which would provide improved screen reader output.

\vspace{-0.2cm}
\paragraph*{Acknowledgments.}
Thanks to Deyan Ginev, Michael Kohlhase, Peter Krautz\-berger, Bruce Miller and Volker Sorge for fruitful discussions.
Thanks also to Fr\'{e}d\'{e}ric Wang for help with the code review. 
A special thanks to Howard Cohl for his editorial work on this paper.
M. Schubotz would like to express his gratitude to Terry Chay and Matthew Flaschen, 
for their help in organizing his internship at the Wikimedia Foundation.
This work was funded by MediaBotz and the Wikimedia Foundation.
\vspace{-0.2cm}

\printbibliography
\end{document}